 \documentclass[final,3p, times,numbers]{elsarticle}

\usepackage{amssymb}
\usepackage{lipsum}

\usepackage{amsmath,amsfonts,amssymb,bm,ulem,cuted}
\usepackage{graphicx}
\usepackage{color}

\newcommand{\red}[1]{{\color{red}{#1}}}

\usepackage[colorlinks=true, pdfstartview=FitV, linkcolor=purple, citecolor= purple, urlcolor=blue]{hyperref}

\journal{Physics Letters B}

\begin{document}

\begin{frontmatter}



\title{The spin alignment of vector mesons with  light front quarks}

\author[label1,label2]{Baochi Fu}
\ead{fubaochi@pku.edu.cn}

\author[label3] {Fei Gao\corref{cor1}}
\ead{fei.gao@bit.edu.cn}\cortext[cor1]{corresponding author}

\author[label1,label2]{Yu-{X}in Liu}
\ead{yxliu@pku.edu.cn}

\author[label1,label2]{Huichao Song}
\ead{huichaosong@pku.edu.cn}

\affiliation[label1]{organization={Center for High Energy Physics, Peking University},
             city={Beijing},
             postcode={100871},
             country={China}}

 \affiliation[label2]{organization={Department of Physics and State
Key Laboratory of Nuclear Physics and Technology, Peking University},
             city={Beijing},
             postcode={100871},
             country={China}}

 \affiliation[label3]{organization={School of Physics, Beijing Institute of Technology},
             city={Beijing},
             postcode={100081},
             country={China}}


\begin{abstract}
The global spin alignment of the vector meson has been observed  in relativistic  heavy ion collisions, but its theoretical origin is still on hot debates. Here we propose to apply the light front framework to explain this phenomenon since the light front form explicitly describes the hadron spin including both the quark spin and  the orbital angular momentum. After applying the light front spinor,  we find that  the spin alignment  in the polarization of vector mesons with $\rho_{00}>1/3$ can be naturally manifested and in particular,   the obtained spin alignment for $\phi$ meson  is in good agreement with the experimental data. This  implies that to explain the spin alignment it is important  to properly include the contribution from the gluon interactions  that are presented in terms of the orbital angular momentum of the hadron bound state.
\end{abstract}




\end{frontmatter}




\section{Introduction}
\label{introduction}

In relativistic heavy ion collisions, the system can carry a large amount of initial angular momentum and polarize the created particles through the vorticity of the quark-gluon plasma(QGP)~\cite{Liang:2004ph}.
The global spin polarization of $\Lambda$ and $\overline{\Lambda}$ hyperons has been observed in  Au+Au collisions by the STAR collaboration, whose behavior has been well described by the hydrodynamic and transport model calculations with thermal vorticity polarization~\cite{Becattini:2013fla, Fang:2016vpj, Liu:2020flb, Karpenko:2016jyx, Li:2017slc, Fu:2020oxj}. The mechanism of the shear induced polarization has also been recently
 proposed  to explain   the  local spin polarization of hyperons~\cite{Fu:2021pok,Becattini:2021iol}.
Besides  the widely studied hyperon polarization, a ``twin effect" named spin alignment of vector mesons has also been proposed as a possible observable of the global polarization, where the spin density matrix component $\rho_{00}$ deviates from 1/3~\cite{Liang:2004xn}.
However, an unexpectedly large global spin alignment has been reported recently in the $\phi$ meson measurements at RHIC, which cannot be explained by the conventional mechanisms of spin alignment, including vorticity and electromagnetic field, local vorticity loop, quark fragmentation, helicity polarization, and axial charge current fluctuations~\cite{Liang:2004xn, Yang:2017sdk, Xia:2020tyd, Becattini:2013vja, Sheng:2019kmk, Gao:2021rom, Muller:2021hpe}.

The success in describing  the global spin polarization of $\Lambda$ and $\overline{\Lambda}$ hyperons is largely because the hyperon's polarization is carried solely by the strange quark according to the flavour-spin wave function \cite{Liang:2000gz}.
On the other hand,  the unexpected large spin alignment of the vector meson indicates that the spin recombination of the immediate quark may not be enough to describe the meson case.
The difference between the quark spin and the meson spin is technically because the Lorentz boost always mixes the kinematic part and interaction part of the hadron spin  dynamically in the instant form with the conventional Dirac spinors~\cite{Dashen:1966zz,Melosh:1974cu,Ahluwalia:1993xa,Brodsky:1994fz,Brodsky:2003pw, Brodsky:2000ii, Speranza:2020ilk, Lorce:2018zpf}.
This makes it difficult to describe the spin structure of hadrons  by  quarks in the instant form.
Such a discrepancy has been long noticed in the proton case which is considered as ``the spin puzzle'' of the proton\cite{Ma:1991xq,Brodsky:2000ii,Kuhn:2008sy}. The strongly correlated  gluon interaction contributes a large fraction of the proton spin which is realized in terms of the orbital angular momentum of hadrons.
Similarly, the contribution of the gluon interaction has also been noticed in the spin kinetic theory, where the spin alignment can be explained after including the gluon fluctuations~\cite{Sheng:2022ssd, Sheng:2019kmk}.

Besides the careful inclusion of  gluon states and/or the orbital angular momentum,  one can instead apply the light front framework where the spin of hadrons is explicit in terms of the spin of light front quarks and thus has great advantages  in describing the spin  structure of hadrons~\cite{Brodsky:2003pw,Brodsky:2000ii}.
The light front form is  different from the conventional instant form, as it provides a different expansion scheme and contains the interaction term even in the leading order.
In particular, the light front spinor  consistently includes the orbital angular momentum  in the hadron,  which  then leads to a nontrivial result of the vector meson's spin alignment.
In this paper  we illustrate that the meson global spin alignment is naturally manifested after the application of  the light front spinor for quarks.

\section{Light front Wigner function and spin density matrix}
The spin density matrix of the vector meson is related to the Wigner function of the vector meson, which can be written as:
\begin{eqnarray}
W_V^{\lambda\lambda^\prime}=\int d^4 Y e^{i q\cdot Y} <V^\lambda(\frac{Y}{2}+X) V^{\lambda^\prime}(X-\frac{Y}{2}) >
\end{eqnarray}
where $V^\lambda$ is the vector meson state with $\lambda$ denoting the polarization index of the vector meson.   Without losing generality, the vector meson state can be expanded upon the Fock states as $|V> \sim |...A_\mu \bar{\Psi}\Psi...>$ with  the gluon field $A_\mu $ and the quark field $\Psi$. The vector meson state is then related to the vector current in the bilinear form of the quark field at the leading order.  In the momentum space, one can define the meson wave function as~\cite{Brodsky:1994kf,Maris:1999nt,Choi:1997iq,Ding:2015rkn}:
\begin{eqnarray}\label{eq:vectormeson}
\Psi^V_{\rm WF}(p,p^\prime;P)=\epsilon^{\mu*}_{\,\,\lambda}(P)<\bar{\psi}(p)\gamma_\mu\psi(p^\prime)|V^\lambda(P)>
\end{eqnarray}
where $P$ is the momentum of the meson,  $p$ and $p^\prime$ are the momentum of the antiquark and quark field, and   $\epsilon^\lambda_\mu(P)$ is the polarization vector of the vector meson with $\epsilon^\lambda_\mu(P)\epsilon^{*\mu}_{\,\,\lambda^\prime}(P)=\delta^{\lambda}_{\lambda^\prime}$.  Note that the above formula defines the the  meson wave function stands for the quark distribution inside the meson.
Now we focus on the formula of the wave function in Fock space in terms of  the quark creation/annilation operator $a^{\sigma+}_p, a^{\sigma^\prime}_{p^\prime}$, which is written as:
\begin{eqnarray}
\Psi^V_{\rm WF}(p,p^\prime)
=\epsilon^{*\mu}_{\,\,\lambda}\bar{u}^{\sigma}(p)\Lambda_+\gamma_\mu u^{\sigma^\prime}(p^\prime)\Psi^{\lambda,\sigma\sigma^\prime}_{\rm WF},
\end{eqnarray}

\noindent with the component of polarized  wave function as the  quark distribution inside the polarized meson
\begin{eqnarray}
\Psi^{\lambda,\sigma\sigma^\prime}_{\rm WF}\propto<a^{\sigma+}_p a^{\sigma^\prime}_{p^\prime} |V^\lambda(P)>
\end{eqnarray}
Here, $u^{\sigma^\prime}(p^\prime)$ is the quark spinor and the matrix $\Lambda_+$ is the positive energy projection matrix which is written as $\Lambda_+=\frac{P\!\!\!/+M}{2M}$ in the Dirac representation with the meson mass $M$. The positive energy projection is usually taken in the light front framework as the transformation property for the positive and negative energy is different~\cite{Brodsky:2000ii}, while in the Lorentz covariant framework, the projection does not have a impact.
The term $n_{\lambda^\prime}=\epsilon^{*\mu}_{\,\,\lambda^\prime}\bar{u}^{\sigma}(p)\gamma_\mu u^{\sigma^\prime}(p^\prime)$ is proportional to  the polarization of the vector meson which can be immediately seen if putting the Fock space formula into Eq.~(\ref{eq:vectormeson}). One  has $<n_{\lambda^\prime}|V^\lambda>\sim\delta^{\lambda}_{\lambda^\prime}$, and hence, one can treat $n^\lambda$ as the polarization direction of the vector mesons~\cite{Brodsky:1994kf}.  The complete form for  the spin density matrix of the  vector meson $\rho_{\lambda\lambda^\prime}$ is  then written  as:

\rule{\dimexpr(0.5\textwidth-0.5\columnsep-0.4pt)}{0.4pt}%
\begin{eqnarray}
\rho_{\lambda\lambda^\prime}&=&<V^\lambda(P)V^{\lambda^\prime} (P)>/\sum_{\lambda=0,\pm1}<V^\lambda(P)V^{\lambda} (P)>\,,\\
<V^\lambda(P)V^{\lambda^\prime}(P)>&=&\frac{1}{\mathcal{N}}\int d^3p d^3 p^\prime \delta^4(p+p^\prime-P) \epsilon_{\mu}^{\lambda}\bar{u}^{\sigma_1}(p^\prime)\gamma^\mu u^{\sigma_1^\prime}(p)
 \epsilon_{\nu}^{\lambda^\prime}\bar{u}^{\sigma_2}(p)\gamma^\nu u^{\sigma_2^\prime}(p^\prime) \bar{\Psi}^{V,\sigma_1\sigma_1^\prime}_{\rm WF} (p,p^\prime;P)\Psi^{V,\sigma_2\sigma_2^\prime}_{\rm WF} (p^\prime,p;P).\notag
\end{eqnarray}

Now, we introduce the light front spinor in the Dirac representation  which can be written as~\cite{Lepage:1979za, Lepage:1980fj}:
\begin{equation}\label{eq:spinor}
u^{LF,\sigma}(p)=\frac{1}{\sqrt{2p^+}}
\left[ \begin{array}{cc}
p_0+m+\vec{\sigma}\cdot\vec{p}\sigma^3 \\
(p_0-m)\sigma^3+\vec{\sigma}\cdot\vec{p}
\end{array}
\right ]\chi^\sigma,
\end{equation}
with  $p^+=p_0+p_z$ and $p_0=\sqrt{\vec{p}^2+m^2}$. $\chi^\sigma$ is the two component spinor in  rest frame.
Here we simply set the meson into the rest frame and thus the momentum of the meson can be parameterized as $P_\mu=(M,0,0,0)$. The positive energy projection then becomes $\Lambda_+=\left[ \begin{array}{cc}
1,&0 \\
0,&0
\end{array}\right ]$.
The polarization vector can also be simply expressed as:
\begin{eqnarray}
\epsilon_{\mu}^{0}=(0,0,0,1), \epsilon_{\mu}^{\pm1}=\frac{1}{\sqrt{2}}(0,1,\pm i,0).
\end{eqnarray}
Therefore, in the meson rest frame,   one  has  $\tilde{\sigma}^\lambda=\epsilon_{\mu}^{\lambda} \sigma^\mu$ with $\lambda=1,-1,0$  corresponding to the respective polarization of vector meson and relating to the Pauli matrix as $\tilde{\sigma}^{1}=\frac{1}{\sqrt{2}}(\sigma^1+i\sigma^2), \tilde{\sigma}^{-1} =\frac{1}{\sqrt{2}}(\sigma^1-i\sigma^2), \tilde{\sigma}^{0} =\sigma^3$.

Now  one may have the relation between the light front bilinear vector currents in  the meson rest frame as:

\begin{eqnarray}
u^{LF,\sigma\dagger}(p)\Lambda_+\left[ \begin{array}{cc}
\tilde{\sigma}^\lambda,&0 \\
0,&\tilde{\sigma}^\lambda
\end{array}\right ]u^{LF,\sigma^\prime}(p)= \begin{array}{c}
\mathcal{F}^\lambda_{\,\,\lambda^\prime}\chi^{\sigma\dagger}\tilde{\sigma}^{\lambda^\prime}\chi^{\sigma^\prime}
\end{array}
=\frac{\chi^{\sigma\dagger}}{\sqrt{2p^+}}\left[ \begin{array}{cc}
p_0+m+\sigma^3 \vec{\sigma}\cdot\vec{p}
\end{array}
\right ]\tilde{\sigma}^\lambda\left[ \begin{array}{c}
p^\prime_0+m^\prime+\vec{\sigma}\cdot\vec{p^\prime}\sigma^3
\end{array}
\right ]
\frac{\chi^{\sigma^\prime}}{\sqrt{2p^{\prime+}}}.\notag\\
\end{eqnarray}

One then has the transition matrix as:
\begin{equation}\label{TrMa}
\mathcal{F}^\lambda_{\,\,\lambda^\prime}=\mathrm{Tr}\left[\frac{1}{2\sqrt{p^+p^{\prime+}}}(p_0+m+\sigma^3\vec{\sigma}\cdot\vec{p})\tilde{\sigma}^\lambda(p^\prime_0+m^\prime+\vec{\sigma}\cdot\vec{p^\prime}\sigma^3)
\tilde{\sigma}_{\lambda^\prime}\right].
\end{equation}

The transition matrix is to transform the vector current in the  bilinear form from the quark rest frame into  the light front frame, which contains the orbital angular momentum of hadron with the term $\vec{\sigma}\cdot \vec{p}\sigma^3$~\cite{Brodsky:2003pw}.
The inclusion of this orbital angular momentum indicates that the light front quark is  roughly speaking equivalent to the conventional relativistic quark together with the gluon interaction. Consequently, it is equivalent to consider the polarized quark wave function with gluon interaction which can be achieved by solving the respective bound state equation of quarks or the interaction Hamiltonian equation of quark model.    Note that the interaction does not apparently appear in the light front form as it is moved into the so called ``angular condition"~\cite{Carbonell:1998rj}. The angular condition ensures the interaction absent from the explicit form  but   only into the wave function of the bound state.

Now one could first define the spin density matrix $\rho^0_{\lambda\lambda^\prime}$ with  quarks in the rest frame and  then  apply the  transition matrix to transform it into light front quarks. The transformed matrix  is then able to describe the spin density matrix of vector meson.  With  the two component spinor in the rest frame $\chi^\sigma$ and its bilinear  vector  form,   one can easily see that $\rho^0_{\lambda\lambda^\prime}$ has the equal polarization for three directions with $\rho^0_{11}=\rho^0_{-1-1}=\rho^0_{00}=1/3$.  After then, one has the spin density matrix in the light front as:

\begin{eqnarray}
\rho_{\lambda\lambda^\prime}=\mathcal{T}_{\lambda }^{\,\,i^\prime}\rho^0_{i^\prime j^\prime}\mathcal{T}^{\dagger j^\prime}_{\quad\lambda^\prime},\quad\quad\mathcal{T}^{\lambda}_{\,\,\lambda^\prime}=\sqrt{\mathcal{N}}\mathcal{F}^\lambda_{\,\,\lambda^\prime},
\end{eqnarray}
with $g^2=(p^++m)(p^{\prime+}+m^\prime)$ and the normalization factor
$$\mathcal{N}=12p^+p^{\prime+}(3(g^4+p^4_\perp)-4g^2p^2_\perp+2p^2_\perp(p^+-p^{\prime+})^2)^{-1}$$
to keep $\rho_{11}+\rho_{-1-1}+\rho_{00}=1$. The relation  $p_\perp=-p_\perp^\prime$, $p^2_\perp=p_x^2+p_y^2$ has been applied in the derivation.
The spin density matrix in  the light front can be then directly regarded as the spin density of the vector meson.
We then have the component $\rho_{00}$:

\begin{equation}
\label{eq:rho}
\rho_{00}(p^+,p^{\prime+},p_\perp)
=\frac{(g^2+p^2_\perp)^2+p^2_\perp(p^+-p^{\prime+})^2}{3(g^2+p_\perp^2)^2-4g^2p^2_\perp+2p^2_\perp(p^+-p^{\prime+})^2}.
\end{equation}

Note that the light front spinor is essential to describe the polarization of the bilinear form with interaction, and Eq.~\ref{eq:rho} is its natural result. One can easily see that $\rho_{00}$ is always larger than 1/3 for any momentum in Eq.~\ref{eq:rho}, and hence,  the light front spinor naturally induces the spin alignment of vector meson with correct sign and magnitude.

For comparison, one may also repeat the computation in the instant form spinor with the conventional Lorentz transformation in the Dirac representation as:
\begin{equation}
u^{\sigma}(p)=\frac{(p\!\!\!/+m)}{\sqrt{2m(m+p_0)}}
\left[ \begin{array}{cc}
\chi^\sigma \\
\xi^\sigma
\end{array}
\right ],
\end{equation}
and the corresponding transition matrix for the vector is:
\begin{equation}
\mathcal{F}^{i}_{j}=\mathrm{Tr}\left[(p\!\!\!/+m)\epsilon_{\mu}^{i}\gamma_0\gamma^\mu (p\!\!\!/+m)
\epsilon^{*\nu}_{j}\gamma_0\gamma_\nu\right],
\end{equation}
and thus one has $\rho_{00}=\rho_{11}=\rho_{-1-1}=1/3$ for any conventional Dirac spinor in the bilinear form after integrating out of the quark momentum.

 By comparing these two spinor forms, we emphasize that the light front approach defines an intrinsic spin alignment along the quatisation direction as $\tilde{\sigma}^0$-direction by considering the inside orbital angular momentum of the hadron bound state.
In the phenomenological study, such orbital angular momentum will be  polarized through the coupling between the inside angular momentum and the global angular momentum of vorticity. More specifically, the vorticity not only polarizes the spin of quarks but also polarizes the orbital angular momentum of the bound state (or equivalently considered as the gluon spin polarization) and finally leads to the  global spin alignment of the vector meson. This also naturally defines the polarization direction as the direction of the global angular momentum of vorticity which, on average, points to the out-of-plane direction ($-y$ axis) in non-central heavy-ion collisions.
\begin{figure}[t]
\centering\includegraphics[width=0.5\textwidth]{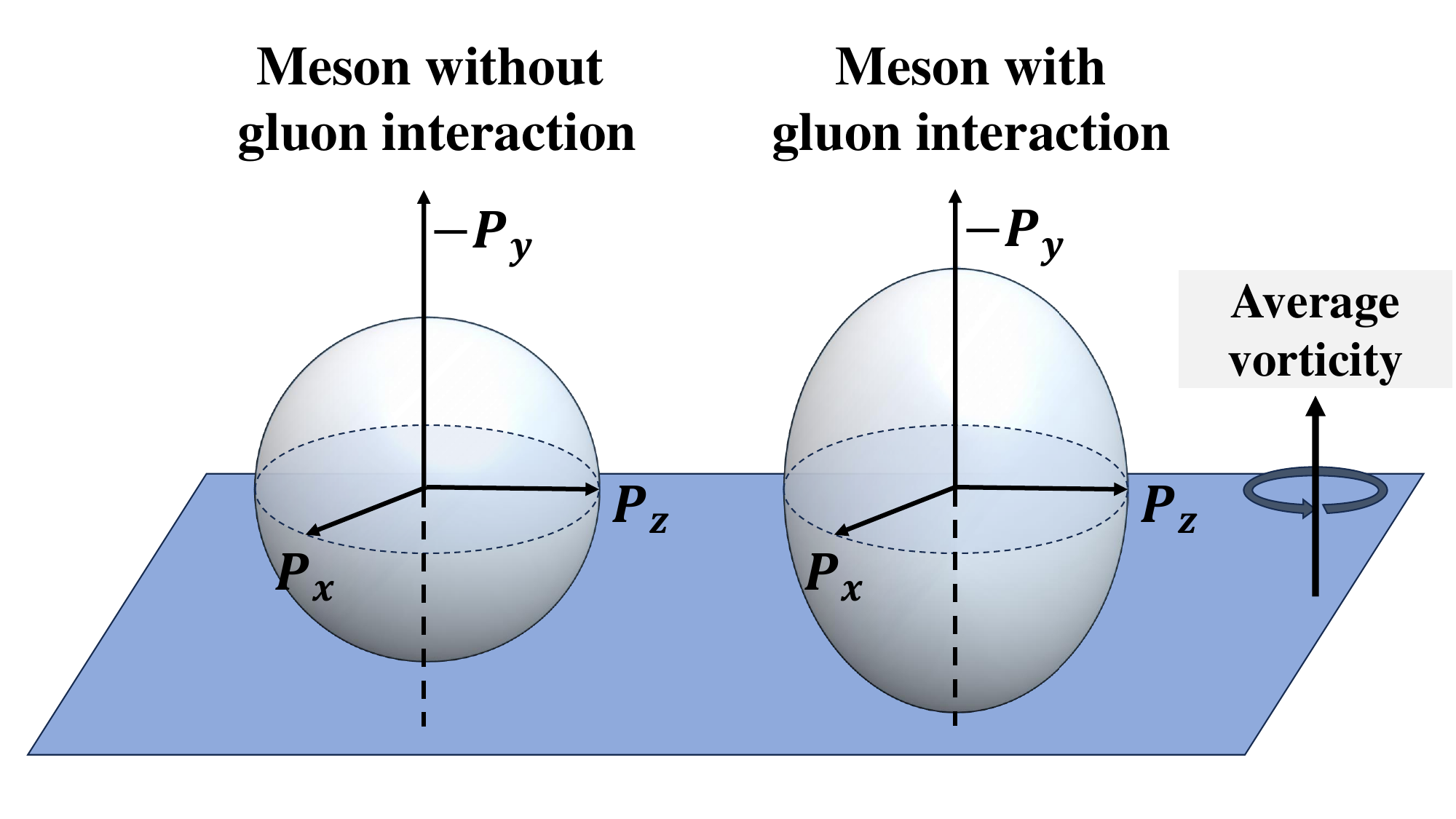}
\caption{A sketch of the intrinsic polarization distribution in the instant form and light front form which contains the internal angular momentum or equivalently, the  gluon interaction.\label{fig:sketch}}
\end{figure}

To illustrate the difference of these two spinor forms, we make a sketch plot in Fig.~\ref{fig:sketch}. In the instant form, the spin of a meson consist of unpolarized quarks is equally distributed and behaves as a sphere structure when the external effect is weak. While in the light front form, the gluon interaction is incorporated in terms of the inside orbital angular momentum of the meson and consequently, even without any external fields, the polarization vector is preferred to along the direction of inside angular momentum as the $\tilde{\sigma}_0$ quantisation direction and shows an ellipsoid distribution (intrinsic spin alignment) in meson rest frame.
Now for a single meson, this quantisation direction is arbitrary and distributed randomly, which means $\rho_{00}=1/3$ when without external angular momentum. However, in the non-central heavy ion collisions with external angular momentum, the inside orbital angular momentum of any particle system will be preferred to direct along the same direction of the average vorticity along $-y$ axis and allows a non-zero average effect (global spin alignment). More specifically, the vorticity not only polarizes the spin of quarks but also polarizes the orbital angular momentum of the bound state and finally leads to the  global spin alignment of the vector meson.

However, to build an exact relation  between the $\tilde{\sigma}_0$-quantisation direction  and the out-of-plane  direction, one  needs  the statistical polarization distribution of  the mesons under the  global polarization of vorticity which can be obtained from the hydrodynamics simulations.
Besides, the full consideration of $P_T$ dependence is also  required which needs to change the definition of the momentum $P$ together the energy and polarization projection and also the parametrization of the quark momentum. These details will be further investigated in the future.  To connect to the  measurements, the momentum distribution is required and $\rho_{00}(p^+,p^{\prime+},p_\perp)$  should be integrated out upon the momentum distribution as will discuss below. However, the difference of momentum distribution will not  change the results qualitatively since despite of these details, the spin density matrix has $\rho_{00}(p^+,p^{\prime+},p_\perp)>1/3$ for any momentum as can be seen from Eq.~\ref{eq:rho}.

\section{Incorporation with the light front wave function}

In the light front framework, one has  $p^+=x P^+,\,p^{\prime+}=(1-x)P^+$, and  $P^+=2\sqrt{m^2+\mu^2}$ with $m$ the respective quark's mass and $\mu$ the  renormalization scale of the quark  in hadron, which is typically chosen as $\mu=1$ GeV~\cite{Ball:1996tb,Arthur:2010xf,Gao:2014bca,Ahmady:2016ujw}. Note that $\mu$ is not a free parameter, instead, it represents the interaction scale of the quarks at the hadronization moment which accounts for the amount of gluon interaction in mesons. This scale gives a microscopic description for the parameters in  the phenomenological studies with the $\phi$ field~\cite{Sheng:2019kmk,Sheng:2022wsy}. It needs to mention that the expression of the light front spinor  shows that   the spin distribution is  more sensitive to the renormalization scale than the momentum distribution as studied in  the unpolarized wave functions~\cite{Ball:1996tb,Arthur:2010xf,Gao:2014bca,Ahmady:2016ujw}, which needs to be carefully considered in  the studies of the polarized wave functions.

Before going to the numerical computations, we firstly  make an estimation in the equal quark mass case with $x=1/2$ and neglecting the mass of quarks and hadrons.
The relation between $\rho_{00}$ and the meson's energy $P^+$ and the quark transversal momentum $p_\perp$ then becomes:
\begin{equation}
{\rho}_{00}=\frac{1}{3-(p_\perp/P^+)^2/((p_\perp/P^+)^2+\frac{1}{4})^2},
\end{equation}
In this case, $\rho_{00}$ is always larger than 1/3, and reaches the limit  $\rho_{00}=1/3$ with $P^+>> p_\perp$. A sketch plot is shown in Fig.~\ref{fig:ptrho} with $p_\perp=0.5$ GeV. $\rho_{00}$  gradually decreases from $0.43$ to the limit 1/3 as $P^+$ goes to infinity.

For a complete calculation of the spin density matrix, one needs to integrate out of the momentum distribution of the quark.
For the thermalized case at finite temperature, a full knowledge of the thermal distribution  of quarks is needed~\cite{Li:2022vmb, Wagner:2022gza}, which is out of current scope and will be further investigated in the future.
For simplicity, we assume that the quark momentum distribution   at hadronization temperature  is not the thermal Fermi-Dirac distribution. Instead, we take the light front wave function $\psi(x,p_\perp)$ in this work. The difference is that the thermal distribution does not contain the gluon interaction which will shift the momentum distribution.  One may apply the coalescence model with considering the angular momentum  between quarks, which will be investigated further. Nevertheless,  the momentum distribution does not affect our results as for any momentum, the above formulae gives $\rho_{00}>1/3$. For the temperature correction on the light front wave function,
We put it in  by modifying the energy scale  as $P^+=2\sqrt{m^2+\mu^2+(2\pi T_{\rm eff})^2}$ with $m$ and $\mu$ the same as the vacuum case above and additionally, a parameterized effective temperature $T_{\rm eff} \sim \tau_0^{-1/3} (dN_{\rm ch} / d\eta)_{\eta = 0}^{1/3}$~\cite{Sheng:2019kmk} is applied with $T_{\rm eff} = 300$~MeV at $\sqrt{s_{NN}} = 200$~GeV:
\begin{equation}
T_{\rm eff} \sim (1/\sqrt{s})^{-1/3} \times (-0.4 + 0.39 \ln{s})^{1/3} .
\end{equation}
It is  also interesting to consider a case that one quark  in the meson is fully thermalized with $P^+\sim T_{\rm eff}>> p_\perp$, then the transition matrix  for this quark becomes unity and Eq.~(\ref{TrMa}) now becomes:
\begin{equation}
\mathcal{F}^{i}_{j}=\mathrm{Tr}\left[\frac{1}{\sqrt{2p^{\prime+}}}\tilde\sigma^i(p^\prime_0+m^\prime+\vec{\sigma}\cdot\vec{p^\prime}\sigma^3)
\tilde\sigma_j\right],
\end{equation}
{This means that regardless of whether the other quark is thermalized or not, the spin density matrix has $\rho_{00}=1/3$, which could partly explains the $\rho_{00}$ difference between $\phi$ and $K^{*0}$. A similar behavior has been mentioned in \cite{Liang:2004xn}, where the coalescence between a polarized quark and an un-polarized quark leads to $\rho_{00} = 1/3$.

\begin{figure}[hbt]
\centering
\includegraphics[width=0.5\textwidth]{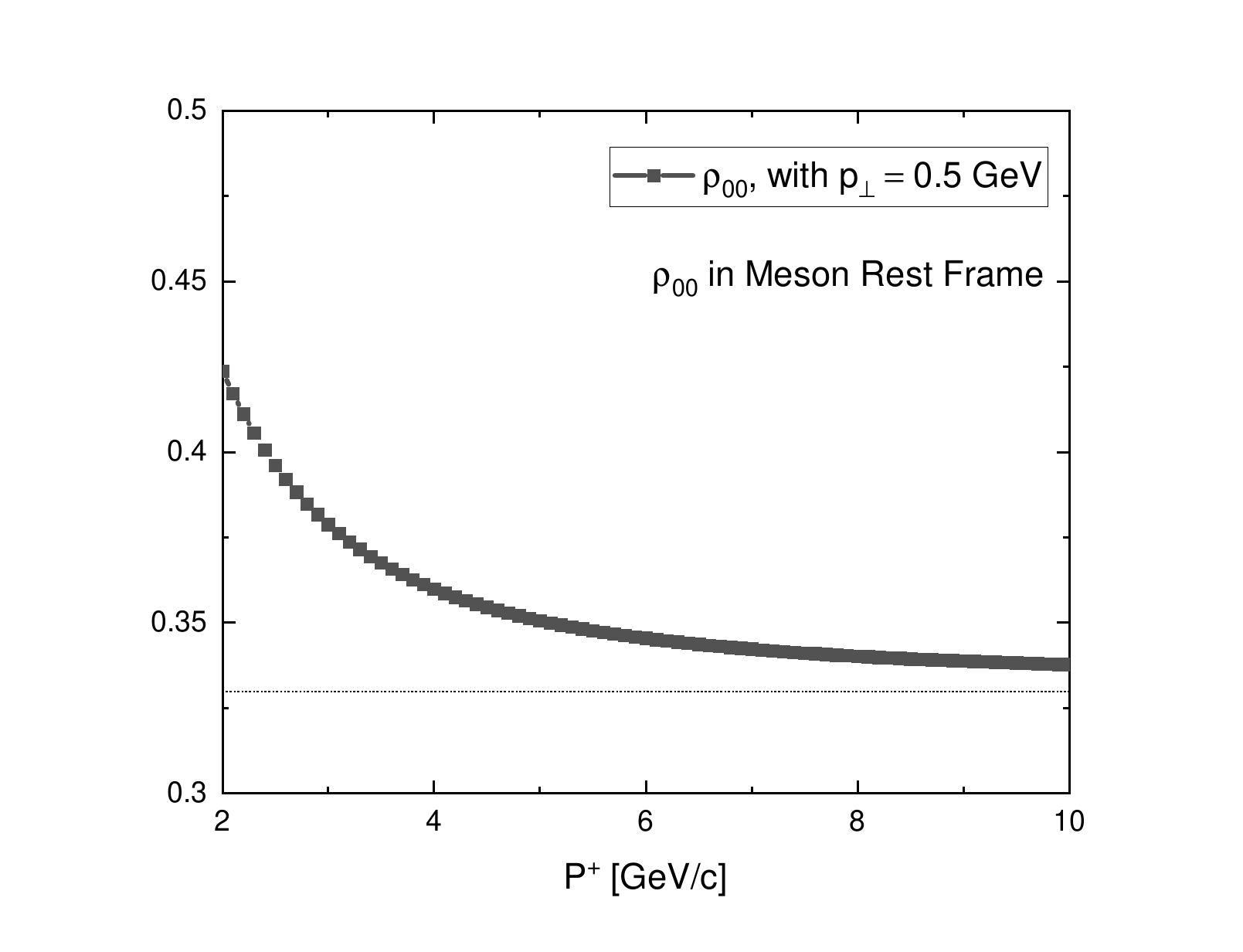}
\caption{A sketch of $P^+$ dependence of $\rho_{00}$ with setting $x=1/2$ and $p_\perp=0.5$ GeV.\label{fig:ptrho}}
\end{figure}

\begin{figure}[t]
\centering
\includegraphics[trim={0.5cm 0cm 1cm 0cm},clip,width=0.9\textwidth]
{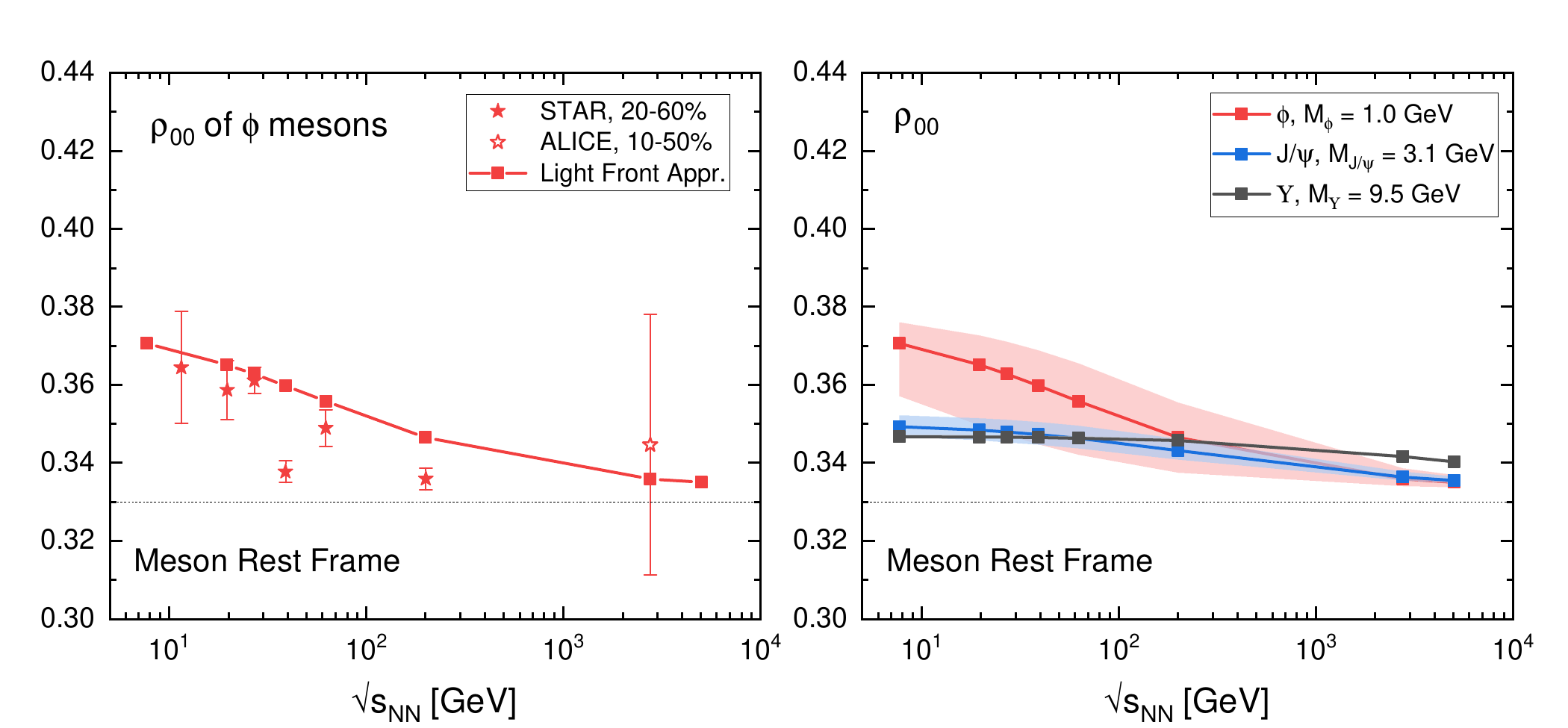}
\caption{Collision energy dependence of $\rho_{00}$ after integrating out of the momentum distribution in the meson rest frame (right panel).   As a comparison, the left panel shows the experimental data of $\phi$ meson from STAR and ALICE at the out-of-plane direction (-y axis)~\cite{STAR:2022fan, ALICE:2019aid}, together with our result for $\phi$ meson which is in good agreement with the measurements. In the right panel, the band shows the corresponding range of $\rho_{00}$ with varying the renormalization scale $\mu$ by $\pm 10 \%$ for each meson. \label{fig:ed}}
\end{figure}


As explained above, we simply apply the light front wave function for $\phi$, $J/\Psi$  and $\Upsilon$ meson in this paper. Then, the average $\rho_{00}$ can be directly defined as:
\begin{eqnarray}\label{eq:averrho}
\bar{\rho}_{00}=\int \frac{dxd^2p_\perp}{x(1-x)} \rho_{00}(x,p_\perp)|\psi(x,p_\perp)|^2.
\end{eqnarray}
The associated wave function $\psi(x,p_\perp)$ can be estimated by separating the momentum distribution wave function in Eq.~(\ref{eq:averrho}), and  parameterized with the Brodsky-Huang-Lepage (BHL) prescription  ~\cite{Huang:1994dy,Braguta:2006wr,Wu:2013lga,Leitao:2017esb} as:
\begin{eqnarray}
\psi(x,p_\perp)=\mathcal{A}\sqrt{x(1-x)} e^{-\frac{p^2_\perp/ (m^2+\mu^2)+\kappa^2}{x(1-x)}},
\end{eqnarray}
where $M$ is the respective meson mass, and $\mathcal{A}$ is the normalization factor for the parton distribution function $q(x)$, defined as:
\begin{eqnarray}
q(x)=\frac{1}{x(1-x)}\int d^2p_\perp |\psi(x,p_\perp)|^2, \  \  \int dx q(x)=1.\notag\\
\end{eqnarray}

In the chiral limit with $g^2=x(1-x)(P^+)^2$,  the above BHL prescription leads to a simple relation between the parton distribution function and the decay constant since the $k_\perp$ dependence can be fully scaled as $k^2_\perp/x(1-x)$ and the $x$ dependence is the same for the transversally and longitudinally polarized vector meson with the longitudinal direction being the spin quantization $\sigma_3$-direction in the rest frame.
One can then estimate the spin density matrix element  with the transversal and longitudinal decay constants of the vector meson at the leading twist~\cite{Ball:1996tb,Gao:2014bca,Ding:2015rkn}, $f_V$ and $f^T_V$, and  one has approximately $\bar{\rho}_{00}\approx\frac{f^2_V}{f^2_V+2(f^T_V)^2}$ in the meson rest frame. With the large $N_c$ limit result of $f^T_V/f_V=1/\sqrt{2}$~\cite{Cata:2008zc,Jansen:2009hr},  an upper limit for the spin alignment may also exist as $\bar{\rho}_{00}=1/2$.

Note that the parameter $\kappa$ mainly influences the shape of the wave function and can be determined by the parton distribution amplitudes~\cite{Braguta:2007fh,Choi:2007ze,Ding:2015rkn}. Here the amplitude can be  expressed as:
\begin{eqnarray}
\phi(x)&=&\frac{1}{\sqrt{x(1-x)}}\int d^2p_\perp \psi(x,p_\perp)\notag\\
&=&\mathcal{A}x(1-x) e^{-\frac{\kappa^2}{x(1-x)}}.
\end{eqnarray}

\begin{table}
\centering
\begin{tabular}{l c c c}
 \hline
 & $\phi$& $J/\Psi$& $\Upsilon$\\
 \hline
  $\kappa$ & $0$ & $1.5$&  $3.0$  \\
  $m$[GeV] & $0.1$ & $1.2$ & $4.2$ \\
  $M$[GeV] & $1.0$ & $3.1$ & $9.5$ \\
 \hline
\end{tabular}
\caption{Parameter $\kappa$ of the light front wave function for $\phi$, $J/\Psi$ and $\Upsilon$ meson and the meson masses $M$ together with the respective  current quark masses  $m$ for $s$, $c$ and $b$ quark.
}
\label{tab:param_m}
\end{table}

We then  perform a numerical calculation, with  parameter $\kappa$  matching the  parton distribution amplitudes in the early studies~\cite{Ding:2015rkn}.  Tab.~\ref{tab:param_m} shows the choice of the parameter  together with the meson mass and the respective current quark masses.
Fig.~\ref{fig:ed} clearly shows that for all three types of mesons, the integrated matrix $\bar{\rho}_{00}$  is larger than $1/3$ at lower collision energies, even for $J/\Psi$ and $\Upsilon$ mesons where the $\rho_{00}$ is suppressed by heavier mass.
Moreover, for larger collision energies, the  increased temperature suppresses the interaction and the related intrinsic spin alignment. The collision energy dependence of the obtained spin alignment for $\phi$ meson  is in good agreement with the experimental data~\cite{STAR:2022fan, ALICE:2019aid, ALICE:2022dyy}.

Note that the results are obtained without fully considering the thermal effect on quark distribution and the effects of bulk evolution. These effects, together with the model uncertainties, can be roughly taken into account by varying the interaction scale $\mu$ in the light front form. In the right panel of Fig.~\ref{fig:ed}, the band shows the possible range of $\rho_{00}$ when the the scale $\mu$ is varied by 10\% percent, which dose not change the qualitative feature and always describes the experimental data within the error bars.
However, one needs the knowledge from the kinetic theory and/or hydrodynamics  to understand the quark distribution  and its evolution,  as well as the time and pattern of hadronization to get  a more descent estimate of the model uncertainties.
For future research, we plan to develop the conventional coalescence model with light front spinor incorporating with hydrodynamics to make a thorough investigation for the global spin alignment of the vector mesons including $K^*$, $J/\Psi$ and $\Upsilon$.

\section{Summary}

To  understand the polarization of the hadrons, one may first separate the effect from the spin distribution and the momentum distribution.  This can be understood within the Fock representation. For instance, for the quark field, one has $\psi(p)\propto u(p) a^\dagger_{\vec{p}}$ with operators like $a_{\vec{p}}$  determining the momentum distribution and the spinors like $u(p) $ determining the spin distribution.

In this work, we emphasize the importance of the light front spinor to deal with the spin alignment of vector mesons. As already mentioned above, the Lorentz boost will mix the kinematic part and the interaction part of hadron's spin  with the conventional Dirac spinor. The conventional way to sum up the spins of quarks and antiquarks in the instant form is not complete to describe the spin of hadrons.  The contribution from the gluon interactions are presented in terms of the orbital angular momentum of the hadron bound state.

Here, instead of involving the interaction term of gluons or considering carefully about its mixture with the angular momentum, we apply the light front spinor of quarks which is explicit in the hadron's spin and contains the angular momentum consistently. The light front form then naturally induces the spin alignment in the polarization of vector mesons with $\rho_{00}>1/3$. Note that this result is independent of the modelling of the  momentum distribution in hadron since it holds for any momentum.
The magnitude of $\rho_{00}$ excess relies on the renormalization scale $\mu$, which is not a free parameter but depends on the interaction scale of the quarks at the hadronization moment. The fixing of the scale requires the knowledge of evolution procedure and the generating mechanism of the respective mesons, which requires further investigations together with hydrodynamic simulation and/or chiral kinetic theories. Nevertheless,  the description we proposed here offers a dynamical mechanism of the spin alignment which connects global spin properties of the hadron  with its inside structure.

After applying a simple ansatz for the wave function of the vector meson, we show that the value of $\rho_{00}$ can be approximately related to the ratio of the transversal and longitudinal decay constants.
Further, we apply the vacuum light front wave function for various mesons and find obvious spin alignment of $\phi$, $J/\Psi$ and $\Upsilon$ gives $\rho_{00} > 1/3$. The spin alignment increases with the decreasing collision energy which is consistent with the experimental measurements of $\phi$ meson qualitatively.
However, to fully describe the polarization and spin alignment of hadrons in heavy ion collisions, the thermalized distribution of quarks and the hadronization procedure are needed to be considered.  We expect to incorporate the kinetic theory and hydrodynamics to further study this in the future.

\section*{Acknowledgments}

We thank Yi Yin, Qun Wang,  Shi Pu and Xinli Sheng for insightful discussions. FG also thanks the other members of the  fQCD collaboration \cite{fQCD}  for discussions and collaboration on related subjects.  YL and HS are supported by  the National  Science Foundation of China under Grants  No. 12247107, No. 12175007 and No.~12075007,
  BF is supported by the National Natural Science Foundation of China under Grants No. 12147173.



\biboptions{sort&compress}
\bibliographystyle{elsarticle-num}

\end{document}